\documentclass[conference]{IEEEtran}
\IEEEoverridecommandlockouts

\usepackage[T1]{fontenc}
\usepackage[utf8x]{inputenc} 

\usepackage{jabbrv}
\usepackage[numbers, sort&compress, square]{natbib}

\DefineJournalException{Proceedings of the 2015 {{IEEE International Conference}} on {{Acoustics}}, {{Speech}} and {{Signal Processing}}}{Proc. ICASSP}

\DefineJournalException{Proceedings of the 2023 {{IEEE International Conference}} on {{Acoustics}}, {{Speech}}, and {{Signal Processing}}}{Proc. ICASSP}

\DefineJournalException{Proceedings of the {{IEEE International Conference}} on {{Acoustics}}, {{Speech}} and {{Signal Processing}}}{Proc. ICASSP}

\DefineJournalException{EURASIP Journal on Audio, Speech, and Music Processing}{EURASIP JASM}

\DefineJournalException{Transactions of the International Society for Music Information Retrieval}{Trans. ISMIR}

\DefineJournalException{Proceedings of the 18th {{International Society}} for {{Music Information Retrieval Conference}}}{Proc. ISMIR}

\DefineJournalException{IEEE/ACM Transactions on Audio, Speech, and Language Processing}{IEEE/ACM TASLP}

\DefineJournalException{IEEE Open Journal of Signal Processing}{IEEE OJSP}

\DefineJournalException{IEEE Open J. Signal Process.}{IEEE OJSP}

\DefineJournalException{Proceedings of the {{International Conference}} on {{Language Technologies}} for {{All}}}{Proc. LT4All}

\DefineJournalException{Proceedings of the {{Twelfth Language Resources}} and {{Evaluation Conference}}}{Proc. LREC}
\DefineJournalException{Proceedings of the {{Thirteenth Language Resources}} and {{Evaluation Conference}}}{Proc. LREC}

\DefineJournalException{Proceedings of the 2024 {{Joint International Conference}} on {{Computational Linguistics}}, {{Language Resources}} and {{Evaluation}} ({{LREC-COLING}} 2024)}{Proc. LREC-COLING}

\DefineJournalException{Proceedings of the 44th {{German Conference}} on {{Artificial Intelligence}}}{Proc. German Conf. AI}

\DefineJournalException{Proceedings of the 6th {{Workshop}} on {{Spoken Language Technologies}} for {{Under-Resourced Languages}} ({{SLTU}} 2018)}{Proc. SLTU}

\DefineJournalException{Proceedings of the 1st {{Joint Workshop}} on {{Spoken Language Technologies}} for {{Under-resourced}} Languages ({{SLTU}}) and {{Collaboration}} and {{Computing}} for {{Under-Resourced Languages}} ({{CCURL}})}{Proc. SLTU/CCURL}

\DefineJournalException{Proceedings of the 2021 Interspeech}{Proc. Interspeech}
\DefineJournalException{Interspeech}{Proc. Interspeech}
\DefineJournalException{Proceedings of the {{Interspeech}} 2017}{Proc. Interspeech}
\DefineJournalException{Proc. {{Interspeech}} 2020}{Proc. Interspeech}
\DefineJournalException{Journal of the International Phonetic Association}{J. Int. Phonetic Assoc.}

\DefineJournalException{Proceedings of the 150th {{Audio Engineering Society Convention}}}{Proc. AES Conv.}

\DefineJournalException{Proceedings of the 1st {{International Conference}} on {{Speech Prosody}}}{Proc. Int. Conf. Speech Prosody}

\DefineJournalException{Ethnologue: {{Languages}} of the {{World}}}{Ethnologue: Languages of the World}

\DefineJournalException{Proceedings of the 2019 {{IEEE Workshop}} on {{Applications}} of {{Signal Processing}} to {{Audio}} and {{Acoustics}}}{Proc. WASPAA}

\DefineJournalException{Proceedings of the 5th {{Workshop}} on {{Detection}} and {{Classification}} of {{Acoustic Scenes}} and {{Events}}}{Proc. DCASE Workshop}

\DefineJournalException{Proceedings of the 2019 {{International Conference}} on {{Learning Representations}}}{Proc. ICLR}

\DefineJournalException{Proceedings of the 125th {{Audio Engineering Society Convention}}}{Proc. AES Conv.}

\DefineJournalException{Proceedings of the 154th {{Audio Engineering Society Convention}}}{Proc. AES Conv.}

\DefineJournalException{Proceedings of the 2022 {{IEEE International Conference}} on {{Acoustics}}, {{Speech}}, and {{Signal Processing}}}{Proc. ICASSP}

\def\authorname{Karn N. Watcharasupat, Chih-Wei Wu, and Iroro Orife}
\def\papersubject{Remastering Divide and Remaster:
A Cinematic Audio Source Separation Dataset with Multilingual Support}

\usepackage[bookmarks=false,%
    pdfauthor={\authorname},%
    pdfsubject={\papersubject},%
    hidelinks]{hyperref}

\usepackage{amsmath}
\usepackage{amsfonts}
\usepackage{amssymb}
\usepackage{amsthm}
\usepackage{mathtools}

\usepackage{microtype}

\usepackage{paralist}
\usepackage{multirow}

\usepackage{adjustbox}

\usepackage{cleveref}

\usepackage{placeins}

\usepackage[ruled,noend,linesnumbered]{algorithm2e}

\DontPrintSemicolon
\SetKwProg{Init}{init}{}{}
\SetAlCapHSkip{0pt}
\SetKwInput{KwInit}{Init}

\SetArgSty{textnormal}

\usepackage{enumitem}

\usepackage{url}

\usepackage{tabularx}
\usepackage{booktabs}
\usepackage{tablefootnote}
\usepackage{siunitx}
\usepackage{diagbox}

\Crefname{algocf}{Algo.\@}{Algos.\@}
\Crefname{figure}{Fig.\@}{Figs.\@}
\Crefname{table}{Tab.\@}{Tabs.\@}

\renewcommand{\arraystretch}{0.75}

\SetAlCapNameFnt{\small}
\SetAlCapFnt{\small}

\begin{document}
\bstctlcite{IEEE:BSTcontrol}
%
\title{Remastering Divide and Remaster:\\
A Cinematic Audio Source Separation Dataset\\with Multilingual Support}

\author{%
    \IEEEauthorblockN{%
        Karn N. Watcharasupat\IEEEauthorrefmark{1}\textsuperscript{($\star$)}\textsuperscript{,}\IEEEauthorrefmark{7},
        Chih-Wei Wu\IEEEauthorrefmark{1},
        and Iroro Orife\IEEEauthorrefmark{1}, 
    }%
    \IEEEauthorblockA{%
        \IEEEauthorrefmark{1}%
        Audio Algorithms, Netflix, Inc.,
        Los Gatos, CA 95032, USA (\textsuperscript{$\star$}Internship)\\
    }%
    \IEEEauthorblockA{%
        \IEEEauthorrefmark{7}%
        Music Informatics Group,
        Georgia Institute of Technology,
        Atlanta, GA 30332, USA\\
    }%
    \IEEEauthorblockA{%
        Email: kwatcharasupat@gatech.edu, \{chihweiw, iorife\}@netflix.com
    }
}

\maketitle

\begin{abstract}
Cinematic audio source separation (CASS), as a problem of extracting the dialogue, music, and effects stems from their mixture, is a relatively new subtask of audio source separation. To date, only one publicly available dataset exists for CASS, that is, the Divide and Remaster (DnR) dataset, which is currently at version 2. While DnR v2 has been an incredibly useful resource for CASS, several areas of improvement have been identified, particularly through its use in the 2023 Sound Demixing Challenge. In this work, we develop version 3 of the DnR dataset, addressing issues relating to vocal content in non-dialogue stems, loudness distributions, mastering process, and linguistic diversity. In particular, the dialogue stem of DnR v3 includes speech content from more than 30 languages from multiple families including but not limited to the Germanic, Romance, Indo-Aryan, Dravidian, Malayo-Polynesian, and Bantu families. Benchmark results using the Bandit model indicated that training on multilingual data yields significant generalizability to the model even in languages with low data availability. Even in languages with high data availability, the multilingual model often performs on par or better than dedicated models trained on monolingual CASS datasets.
\end{abstract}


%
\IEEEpeerreviewmaketitle

\section{Introduction} \label{sec:intro}

Cinematic audio source separation (CASS), as a problem of extracting dialogue (DX), music (MX), and effects (FX) stems from their mixture, is a relatively young subtask of audio source separation. Unlike domain-specific source separation such as speech or music source separation, the audio content tackled by CASS spans essentially all possible natural and artificial sounds that could be recorded, synthesized, or otherwise produced. While CASS shares some similarities with universal audio source separation \cite{Kavalerov2019UniversalSoundSeparation}, CASS deals with audio source separation within an inherently creative domain, with a context-dependent and malleable categorization of sound classes into the three-stem setup. In particular, many musical instruments can be used to create non-musical sound effects. In this context, these sounds from musical instruments should be placed into the FX stem. Similarly, many objects typically not considered musical instruments have often been used in a musical context and their sounds should be placed into the MX stem, as per the contextual usage of the sounds. As expected with artistic uses of sounds, many additional edge cases and variations in practice exist when it comes to the categorization of sound into the DX, MX, and FX stems\footnote{See \href{https://tinyurl.com/nflx-mne-guidelines}{tinyurl.com/nflx-mne-guidelines} for examples.}.

Efforts to improve the perceived sound quality or intelligibility of cinematic audio content date back to at least \cite{Uhle2008SpeechEnhancementMovie}. In the context of streaming media over the internet, CASS can act as a versatile preprocessor that opens up possibilities for audience-side personalization of cinematic content \cite{Paulus2019SourceSeparationEnabling, Rieger2023DialogueEnhancementMPEGH}. The idea of CASS as a three-stem source separation problem was first introduced by P{\'e}termann et al\@. \cite{Petermann2022CocktailForkProblem, Petermann2023TacklingCocktailFork}, together with the Divide and Remaster (DnR) dataset, a synthetic dataset with raw content drawn from LibriSpeech \cite{Panayotov2015LibrispeechASRCorpus}, FMA~\cite{Defferrard2017FMADatasetMusic}, and FSD50K \cite{Fonseca2022FSD50KOpenDataset}.  DnR was then used in the Cinematic Audio Demixing (CDX) track of the 2023 Sound Demixing Challenge \cite{Uhlich2023SoundDemixingChallenge}, with participation from various academic and industry groups. In the process, several areas of improvement to DnR, still the only publicly available CASS dataset, have been suggested. Specifically, Uhlich et al\@.~\cite{Uhlich2023SoundDemixingChallenge} and other researchers have identified potential issues concerning the mismatch in production quality of the data from real cinematic audio, the lack of emotional content of the dialogue data, and vocal contents in non-dialogue stems. Our informal internal listening test of the Bandit model \cite{Watcharasupat2023GeneralizedBandsplitNeural} on real Netflix content also indicated that the model, trained on DnR v2, often struggles with non-English dialogue, confusing unseen phonemes and use of linguistic tones as music or effects.


In this work, we developed version 3 of the DnR dataset\footnote{Landing page: \href{https://github.com/kwatcharasupat/source-separation-landing}{github.com/kwatcharasupat/source-separation-landing}}. While the dataset remains synthetic, it is our hope that version~3 now more closely reflects not just the common but also the diverse variations in cinematic audio production across languages, regions, genres, and creative practices. The major changes from DnR v2 are as follows:
\begin{itemize}[leftmargin=*]
    \item the dialogue stem now contains content from more than 30 languages across various language families;
    \item speech, vocals, and/or vocalizations have been removed from the music and effects stems;
    \item loudness and timing parametrization have been adjusted to approximate the distributions of real cinematic content;  and,
    \item the mastering process now preserves relative loudness between stems and approximates standard industry practices.
\end{itemize}
As a benchmark, we trained Bandit models \cite{Watcharasupat2023GeneralizedBandsplitNeural} on multiple linguistic variants of the proposed dataset and demonstrated that the multilingual model often performs on par or better than dedicated monolingual models on their respective test sets. We have made efforts to ensure that the proposed dataset was derived entirely from contents that permit commercial use, derivative work, and redistribution. The resulting dataset will be released under the CC BY-SA 4.0 License, while the replication code will be released under the Apache 2.0 license. Full attributions will be provided in the repository. The authors hope that the dataset will continue to be improved with more diverse sonic inventories for all three stems as more applicable datasets are developed and/or identified.

\section{Dataset Setup} \label{sec:dataset-setup}

The Divider and Remaster dataset v3 consists of three splits and multiple linguistic variants each: train (6000 clips per variant), validation (600 clips per variant), and test (1200 clips per variant).
Each clip is a collection of DX, MX, foreground FX (FGFX), background FX (BGFX), their mixture, and the combined FX stems. The dialogue stem in each variant draws from a different language or group of languages. For each identically indexed clip across variants, the music and effects stem share identical underlying pre-mastering tracks, but will differ slightly in the final tracks due to the mastering that depends on the mixture loudness.

All audio data are mono and $D_{\text{track}}=\SI{60}{s}$ in duration, as with DnR v2.  However, audio data in v3 are sampled at \SI{48}{\kilo\hertz} with a bit depth of 24 bit, in line with delivery specifications used by major cinematic content providers such as Netflix, Sony Pictures, Warner Bros.\@ Discovery, and Hulu\footnote{See {tinyurl.com/\{\href{https://tinyurl.com/nflx-delivery-specs}{nflx}, \href{https://tinyurl.com/sony-delivery-specs}{sony}, \href{https://tinyurl.com/wbd-delivery-specs}{wbd}, \href{https://tinyurl.com/hulu-delivery-specs}{hulu}\}-delivery-specs}.}.

\subsection{On the Lack of Spatialization}

Although most cinematic content is multichannel, DnR~v3 remains a one-channel dataset. The rationale for this decision is due to the very open problem of \textit{algorithmically} generating realistic source trajectories. Moreover, even though many methods for artificial spatialization exist for arbitrary trajectories and channel layouts, they are often unable to recreate realistic spatialized reverberations without relying on human-in-the-loop artistic intervention. Providing DnR v3 beyond mono without meaningful spatialization would provide little additional benefits in teaching a model to address spatial coherence issues. Therefore, we decided to remain in one channel for the time being. Multichannel inference can still be performed on models trained on mono data by treating each channel as a pseudo-mono track, albeit likely with some phase coherence issues.

\subsection{License Consideration} \label{sec:dataset-setup/license}

Our best efforts have been made to check that the data used in this work permits the creation of derivative works, commercial use, or redistribution of the original/derivative works. For datasets where the licenses were specified at the track level, we have verified each license individually and excluded any file that does not meet the above requirements from further development in this work. Any data files with missing or unclear licenses were also excluded. 

\subsection{Music Data and Preprocessing} \label{sec:dataset-setup/mx}

As with DnR v2, we utilize the {Free Music Archive} (FMA) dataset \cite{Defferrard2017FMADatasetMusic} for the music stem. FMA provided MP3-encoded stereo audio data sampled at \SI{44.1}{\kilo\hertz}. The audio tracks in FMA are individually licensed thus only a subset permits commercial and derivative use. Specifically, only public-domain audio tracks and those licensed under the CC BY, CC BY-SA, Open Audio License, or Free Art License families were included. The distribution of the licenses of FMA audio tracks can be found \Cref{tab:fma}. In total, \num{11175} (\SI{10.5}{\percent}) out of \num{106574} tracks were included for further development.

\begin{table}[t]
    \caption{Distribution of the licenses of the audio tracks in FMA. }
    \setlength{\tabcolsep}{3pt}
    \label{tab:fma}
    \centering
    \begin{tabularx}{\columnwidth}{ccX
    S[table-format=5.0] S[table-format=2.3]}
    \toprule
    Com.? & Deriv.? & License & {Count} & {\%}\\
    \midrule
    Y & Y & CC BY (1.0, 2.0, 2.5, 3.0, 4.0) & 6960 & 6.5 \\
     &  & CC BY-SA (2.0, 2.5, 3.0, 4.0) & 2802 & 2.6 \\
     &  & Public Domain & 1392 & 1.3 \\
     &  & Free Art License & 16 & 0.02 \\
     &  & Open Audio License & 5 & 0.005 \\
    \cmidrule{2-5}
     & N & CC BY-ND (2.0, 2.5, 3.0, 4.0) & 962 & 0.9 \\
    \cmidrule{2-5}
     & ? & CC Sampling+ (1.0)& 14 & 0.01 \\
    \midrule
    N & Y & CC BY-NC (2.0, 2.1, 2.5, 3.0, 4.0) & 8313 & 7.8 \\
     &  & CC BY-NC-SA (2.0, 2.1, 2.5, 3.0, 4.0) & 43512 & 40.8 \\
    \cmidrule{2-5}
     & N & CC BY-NC-ND (2.0, 2.1, 2.5, 3.0, 4.0) & 41869 & 39.3 \\
     &  & Free Music Philosophy & 138 & 0.1 \\
     &  & CopyrightPlus & 23 & 0.02 \\
     &  & FMA Music Sharing & 18 & 0.02 \\
     &  & ideology.de License & 10 & 0.009 \\
    \cmidrule{2-5}
     & ? & CC NC-Sampling+ (1.0) & 19 & 0.02 \\
    \midrule
    ? & ? & Sound Recording Common Law & 428 & 0.4 \\
     &  & Missing license & 88 & 0.08 \\
     &  & Orphan work & 5 & 0.005 \\
    \bottomrule
    \end{tabularx}
    \vspace{1pt}
    
    Com.?: Allows commercial use?; Deriv.?: Allows derivative works?

    \vspace{-3ex}
\end{table}

Due to the significant reduction in the number of available tracks, we utilized the full-track version of FMA instead of the 30-second version used in DnR v2. Segments with speech and/or vocals were removed by running a speech-music activity detection (SMAD) model \cite{Hung2022LargeTVDataset} on the track and keeping only contiguous segments without any speech activation with a minimum duration of \SI{5.0}{\second}. To maximize data diversity in terms of mixing parameters, we treated the left and right channels of the data as two pseudo-independent mono tracks, instead of downmixing them into mono tracks as done in DnR v2. After the preprocessing, we have \num{170386} segments of music data without vocal or speech content, totaling \SI{1143.5}{\hour}.

\subsection{Effects Data and Preprocessing} \label{sec:dataset-setup/fx}

As with DnR v2, we also utilized the \textit{FSD50K} dataset \cite{Fonseca2022FSD50KOpenDataset}. Audio tracks in FSD50K are also individually licensed with either CC0, CC BY, CC BY-NC, or CC Sampling+. Only the CC0 and CC BY licensed tracks were used in v3. After filtering, \num{43379} (\SI{84.7}{\percent}) out of \num{51197} tracks were included.

In order to filter out tracks that may include speech or music content, we first refer to the AudioSet ontology \cite{Gemmeke2017AudioSetOntology} used in FSD50K. Any track under the ``Human Voice'', ``Music'', and ``Sound reproduction'' umbrellas were excluded. Within the ``Human group actions'' umbrella, all tracks except those explicitly labeled as ``Clapping'' or ``Applause'' were excluded. Exclusion criteria take priority in tracks with multiple labels.

Of the \num{25147} remaining tracks, we further ran SMAD on any file longer than \SI{1}{\second} in duration and excluded those with any detected speech or music frame. After the preprocessing, we have \num{21760} segments of non-vocal FX data, totaling \SI{41.3}{\hour}.

Unlike DnR v2, however, any sound from any class can be either foreground or background. The only differences between the FGFX and BGFX stems lie in the loudness and density of sound events.

\subsection{Dialogue Data and Preprocessing} \label{sec:dataset-setup/dx}


\begin{table*}[t]
    \centering
    \caption{List of languages included in DnR v3}\vspace{-5pt}
    \label{tab:lang}
    \footnotesize
    \renewcommand{\arraystretch}{0.7}
    \setlength{\tabcolsep}{4pt}
    \begin{tabularx}{\linewidth}{%
        l@{\hspace{1em}}>{\scshape}X
        l@{\hspace{1em}}>{\scshape}X
        lll
        S[table-format=3.1]
        S[table-format=3.0]@{ }
        S[table-format=3.0]@{ }
        S[table-format=3.0]
        S[table-format=4.0, round-mode=places, round-precision=0, exponent-mode=fixed, fixed-exponent=6, drop-exponent=true]
    }
    \toprule
         & & & & & \multicolumn{2}{c}{Original Format} & & \multicolumn{3}{c}{Voices} &\\
        \cmidrule{6-7}\cmidrule{9-11}
        Language & & Family & & Source Dataset & $f_{\text{s}}$ (kHz) & Codec & {Hrs.} & {F} & {M} & {U} & {L1+L2 (M)}\\
    \midrule
        English, unspecified & eng
            & W\@. Germanic & gmw
                & LibriSpeech \cite{Panayotov2015LibrispeechASRCorpus} & 44.1 & MP3, \char`\~128 kbps & 111.4 & 165 &  166 & & 1515231760 \\
        English, Nigerian & eng 
            & W\@. Germanic & gmw & SLR 70 \cite{Butryna2020GoogleCrowdsourcedSpeech} & 48 & PCM, 16-bit & 5.8 & 19 & 12 \\
        English, Irish & eng 
            & W\@. Germanic & gmw & SLR 83 \cite{Demirsahin2020OpensourceMultispeakerCorpora} & 48 & PCM, 16-bit &  0.7 & & 3 &  \\
        English, Scottish & eng 
            & W\@. Germanic & gmw & SLR 83 \cite{Demirsahin2020OpensourceMultispeakerCorpora} & 48 & PCM, 16-bit &  4.3 & 6 & 11 &  \\
        English, Welsh & eng 
            & W\@. Germanic & gmw & SLR 83 \cite{Demirsahin2020OpensourceMultispeakerCorpora} & 48 & PCM, 16-bit &  5.4 & 8 & 11 &  \\
        English, Midlands & eng 
            & W\@. Germanic & gmw & SLR 83 \cite{Demirsahin2020OpensourceMultispeakerCorpora} & 48 & PCM, 16-bit &  1.2 & 2 & 3 &  \\
        English, N\@. England & eng 
            & W\@. Germanic & gmw & SLR 83 \cite{Demirsahin2020OpensourceMultispeakerCorpora} & 48 & PCM, 16-bit &  5.0 & 5 & 14 &  \\
        English, S\@. England & eng 
            & W\@. Germanic & gmw & SLR 83 \cite{Demirsahin2020OpensourceMultispeakerCorpora} & 48 & PCM, 16-bit &  14.6 & 28 & 29 &  \\
        \midrule
        
        German, Standard & deu
            & W\@. Germanic & gmw 
                & HUI \cite{Puchtler2021HUIAudioCorpusGermanHighQuality} & 44.1 & PCM, 16-bit & 253.8 & 61 & 54 & 3  & 133908920\\
        \midrule
        Faroese & fao & N\@. Germanic & gmq & SLR 125 \cite{Simonsen2022CreatingBasicLanguage} & 48 & PCM, 16-bit & 94.5 & && 369 & \multicolumn{1}{r@{ }}{<1}\\
        \midrule
        French & fra & Romance & roa & Audiocite \cite{Felice2024AudiociteNetLarge} & 44.1, 48 & MP3, 64--320 kbps & 3478.3 & 32 & 22 & 4 & 311575110\\
        Italian & ita & Romance & roa & LaMIT \cite{DiBenedetto2022LaMITDatabaseRead} &44.1 & PCM, 16-bit & 0.8 & 2 & 2 && 66789200\\
        Catalan & cat& Romance & roa & SLR 69 \cite{Kjartansson2020OpenSourceHighQuality} & 48 & PCM, 16-bit & 9.4 &20 & 16 && 9299570 \\
        Galician & glg &Romance & roa & SLR 77 \cite{Kjartansson2020OpenSourceHighQuality} & 48  & PCM, 16-bit  & 10.3 & 34 & 10 &  & 3378300\\
        \midrule
        Spanish, Argentinian & spa 
            & Romance & roa 
                &  SLR 61  \cite{Guevara-Rukoz2020CrowdsourcingLatinAmerican} & 48 & PCM, 16-bit & 8.1 & 31 & 13 && 559520830 \\
        Spanish, Peninsular & spa 
            & Romance & roa 
                &  SLR 61  \cite{Guevara-Rukoz2020CrowdsourcingLatinAmerican}  & 48 & PCM, 16-bit & 0.1 & 1 & 2  \\
        Spanish, Chilean & spa
            & Romance & roa & SLR 71  \cite{Guevara-Rukoz2020CrowdsourcingLatinAmerican} & 48 & PCM, 16-bit & 7.1 & 13 & 18 \\
        Spanish, Colombian & spa 
            & Romance & roa 
                &  SLR 72 \cite{Guevara-Rukoz2020CrowdsourcingLatinAmerican}  & 48 & PCM, 16-bit & 7.6 & 16 & 17 \\
        Spanish, Peruvian & spa 
            & Romance & roa 
                &  SLR 73 \cite{Guevara-Rukoz2020CrowdsourcingLatinAmerican}  & 48 & PCM, 16-bit & 9.2 & 18 & 20 \\
        Spanish, Puerto Rican & spa 
            & Romance & roa 
                &  SLR 74  \cite{Guevara-Rukoz2020CrowdsourcingLatinAmerican}  & 48 & PCM, 16-bit & 1.0 & 5 \\
        Spanish, Venezuelan & spa 
            & Romance & roa 
                & SLR 75  \cite{Guevara-Rukoz2020CrowdsourcingLatinAmerican}  & 48 & PCM, 16-bit & 4.8 &11 & 12 \\
        \midrule
        Basque & eus & Paleo-European & & SLR 76 \cite{Kjartansson2020OpenSourceHighQuality} & 48 & PCM, 16-bit & 13.9 &29&23&& 1086860 \\
        \midrule
        Kazakh & kaz & Turkic  & trk
            & SLR 140 \cite{Kadyrbek2023DevelopmentKazakhSpeech} &44.1, 48 & PCM, 16-bit & 11.5 &&& 22 &16579380 \\
        Ukranian & ukr &  E\@. Slavic & zle & \cite{Smoliakov2023OpenSourceUkrainian} &48 & PCM, 16-bit & 16.6 & 2 & 1 & &38919810\\
        &&&  &&& OPUS, \char`\~128 kbps & 6.3 &  & 1 & \\
        \midrule
        Arabic, Levantine & apc & Semitic & sem & ASC \cite{Halabi2016ModernStandardArabic} & 48 & PCM, 16-bit & 4.1 & & 1 && 53518440 \\
        \midrule
        Bengali & ben & Indo-Aryan & inc &SLR 37 \cite{Sodimana2018StepbyStepProcessBuilding} & 48 & PCM, 16-bit & 5.0 & &  15 && 278184510\\
        Nepali & npi & Indo-Aryan & inc &SLR 43 \cite{Sodimana2018StepbyStepProcessBuilding} & 48 & PCM, 16-bit & 2.8 & 18& && 31979760\\
        Marathi & mar & Indo-Aryan & inc &SLR 64 \cite{He2020OpensourceMultispeakerSpeech} & 48 & PCM, 16-bit & 3.0 & 9 &&& 99262870\\
        Gujarati & guj & Indo-Aryan & inc &SLR 78 \cite{He2020OpensourceMultispeakerSpeech} & 48 & PCM, 16-bit & 7.9 & 18 & 18&& 62673620\\
        \midrule
        Malayalam & mal & Dravidian & dra & SLR 63 \cite{He2020OpensourceMultispeakerSpeech} & 48 & PCM, 16-bit & 5.5 & 24 & 18&& 37738880\\
        Tamil & tam & Dravidian & dra & SLR 65 \cite{He2020OpensourceMultispeakerSpeech} & 48 & PCM, 16-bit & 7.1 & 25 & 25&& 86724670\\
        Telugu & tel & Dravidian & dra & SLR 66 \cite{He2020OpensourceMultispeakerSpeech} & 48 & PCM, 16-bit & 5.7 & 24  & 23&& 95810890\\
        Kannada & kan & Dravidian & dra & SLR 79 \cite{He2020OpensourceMultispeakerSpeech} & 48 & PCM, 16-bit & 8.5 & 23 & 36&& 58706690\\
        \midrule
        Malay & zlm &  Malayo-Polynesian & poz & Malaya Speech \cite{Husein2024MalayaSpeech} & 44.1 & PCM, 16/24-bit & 6.7 & 1 & 1 && 19179470\\
        Javanese & jav & Malayo-Polynesian & poz & SLR 41 \cite{Sodimana2018StepbyStepProcessBuilding} &48 & PCM, 16-bit & 7.0 & 19 & 20 && 68335000 \\
        Sundanese & sun & Malayo-Polynesian & poz & SLR 44 \cite{Sodimana2018StepbyStepProcessBuilding} &48 & PCM, 16-bit & 5.4 & 20 & 21 && 32400000 \\
        \midrule
        Khmer & khm & Mon-Khmer & mkh & SLR 42 \cite{Sodimana2018StepbyStepProcessBuilding} &48 & PCM, 16-bit & 4.0 & &  16 && 17600950\\
        Burmese & mya & Tibeto-Burman & tbq & SLR 80 \cite{Oo2020BurmeseSpeechCorpus}&48 & PCM, 16-bit & 4.1 & 20 & && 43165690\\
        \midrule
        Chinese, Mandarin & cmn
            & Chinese & zhx
                & AISHELL-3 \cite{Shi2021AISHELL3MultiSpeakerMandarin} & 44.1 & PCM, 16-bit & 85.6 & 175 & 43 && 1140361330 \\
        Japanese & jpn & Japonic & jpx & JVNV \cite{Xin2024JVNVCorpusJapanese} & 48 & PCM, 24-bit & 3.9 & 2 & 2 & & 123466720\\
            &&&& PJS \cite{Koguchi2020PJSPhonemebalancedJapanese} & 48 & PCM, 24-bit & 0.6 & &  1 \\
        \midrule
        Afrikaans & afr & W\@. Germanic & gmw & SLR 32 \cite{Niekerk2017RapidDevelopmentTTS}&48 & PCM, 16-bit & 3.3 & & & 9 & 18093000 \\
        Xhosa & xho & Bantu & bnt & SLR 32 \cite{Niekerk2017RapidDevelopmentTTS}&48 & PCM, 16-bit & 3.1 &&& 6 & 19216300\\
        Sesotho & sot & Bantu & bnt & SLR 32 \cite{Niekerk2017RapidDevelopmentTTS}&48 & PCM, 16-bit & 3.2 &&& 6 & 13524700\\
        Setswana & tsn & Bantu & bnt & SLR 32  \cite{Niekerk2017RapidDevelopmentTTS} &48 & PCM, 16-bit & 3.5 &&& 5 & 13745730\\
        Yoruba & yor & Defoid &  & SLR 86 \cite{Gutkin2020DevelopingOpenSourceCorpus} &48 & PCM, 16-bit & 4.0 & 19 & 17 && 47195900 \\
    \bottomrule
    \end{tabularx}
    \smallbreak
    
    Languages are grouped by genealogical proximity and/or geopolitical proximity of their primary region(s) of use or origin.
\end{table*}

The dialogue stem in DnR v3 consists of data from 32 languages across 40 datasets, summarized in \Cref{tab:lang}. The datasets were chosen by first going through all available datasets in OpenSLR that satisfy the quality and license requirements\footnote{\href{https://openslr.org/}{openslr.org.} The latest entry at the time of writing was SLR 151.}. We then perform naive web searches for any dataset representing a language within the 200 most spoken languages, as per the 2023 Ethnologue 200, that satisfies the requirements. 

Given the large variety of data formats, any included audio track is required to have a sampling rate of at least \SI{44.1}{\kilo\hertz}, a bit depth of at least 16 bits for lossless encoding, or a bit rate of at least \SI{64}{kbps} for lossy encoding. Manual spot checks were performed on each dataset. Datasets with existing sentence- or utterance-level segmentation were kept as is. Longer audio files were segmented, with leading and trailing silences removed for each segment. Datasets with existing train-validation-test or equivalent splits were respected regardless of the original distributions. Datasets with only train-test splits or equivalent were split into train-validation-test splits by breaking up the original train split into a train split and validation split at speaker level. Any dataset with only one speaker was treated as a train-only dataset. Any dataset with between two to six speakers was split, if possible, to have the test set consist of at least one male and one female speaker, followed by the train and then validation sets, in this order of priority. Dataset-specific details along with linguistic background are detailed below. Linguistic and demographic information was drawn from Ethnologue \cite{Eberhard2024EthnologueLanguagesWorld} unless explicitly cited. All included Google datasets (SLR~32, 37, 41--44, 61, 63--66, 69--80, 83, 86) are licensed under CC BY-SA 4.0. 


\label{sec:dx/eng}

\textbf{English} (\textsc{eng}) is a West Germanic language with more than \SI{1.5}{B} total speakers across the world, making it the most spoken language in the world in terms of total speakers\footnote{``Total speakers'' refers to the total number of ``first-language'' (L1) and ``second-language'' (L2) speakers.}. Due to its widespread use throughout the world, significant regional varieties of English have emerged, including many dialects, creoles, and pidgins, with significant variations in phonology. In general, English is a non-tonal and stress-timed language, with 24 consonants, 13 vowels, and 8 diphthongs.
English data were drawn from LibriSpeech (SLR~12), SLR~70, and SLR~83. 
LibriSpeech \cite{Panayotov2015LibrispeechASRCorpus} is a large-scale dataset of read English audiobook speech drawn from the {LibriVox} project. As with DnR v2, the original \SI{44.1}{\kilo\hertz} MP3 files from LibriVox were used for the creation of DnR v3. LibriSpeech is licensed under CC BY 4.0 while the original LibriVox data is public domain. 
SLR~70 is a dataset of Nigerian English, without pidgin, recorded in Lagos, Nigeria, and London, UK \cite{Butryna2020GoogleCrowdsourcedSpeech}. Based on British English, Nigerian English contains prosodic features closer to a tonal language \cite{Gut2002ProsodyNigerianEnglish}. Note, however, that Nigerian English consists of multiple regional varieties with differing distinctive features. 
SLR~83 is a dataset of various English dialects across the British Isles, namely Irish, Scottish, Welsh, Midlands, Northern England, and Southern England. 

\textbf{Standard German} (\textsc{deu}) is also a West Germanic language with more than \SI{130}{M} total speakers, mainly in Western and Central Europe. German is a non-tonal and stress-timed language, with 22 consonants, 22 vowels, and 3 diphthongs, notably with the rare voiceless labiodental affricate /pf/.  German data were drawn from the {HUI Audio Corpus} \cite{Puchtler2021HUIAudioCorpusGermanHighQuality} which was also derived from LibriVox. The dataset, with almost 120 speakers, is licensed under CC0. Although unstated in \cite{Puchtler2021HUIAudioCorpusGermanHighQuality}, these audio were likely originally obtained from LibriVox as lossily encoded MP3 files with a bit rate of \SI{128}{kbps}.   

\textbf{Faroese} (\textsc{fao}) is a North Germanic language with \SI{69}{K} total speakers, mostly in the Faroe Islands. Descended from Old Norse (\textsc{non}), Faroese has around 20 consonants, more than 20 vowels in some analyses, and 8 diphthongs \cite{Barnes1995Faroese}. The dataset for Faroese is the {Basic Language Resource Kit (BLARK) for Faroese} (SLR~125) \cite{Simonsen2022CreatingBasicLanguage}, licensed under CC BY 4.0. Faroese data were provided unsegmented. Pydub was used to segment the Faroese data into nonsilent segments.

\textbf{French} (\textsc{fra}) is a Romance language, specifically from the Gallo-Romance branch, with more than \SI{300}{M} total speakers across all continents. As with all Romance languages, French descended from Vulgar Latin. It is a non-tonal language with syllable-timed stress, consisting of 20 consonants and 14 vowels. French features guttural R, nasal vowels, and liaison between words.
French data were drawn from the {AudioCité} corpus (SLR~139)
\cite{Felice2024AudiociteNetLarge}, an extremely large read speech corpus with more than 6700 hours of data. AudioCité data are individually licensed and provided in an MP3 format at various bit rates and sampling rates. As a result, we only included sufficiently high-quality data that are public domain, or licensed under CC BY, CC BY-SA, or License Art Libre. {AudioCité} data were provided unsegmented, with some sections containing non-vocal content such as introductory music. SMAD was used to detect non-silent speech segments of at least \SI{1}{\second} in duration and to remove any segment with music. After preprocessing, \SI{1.3}{M} segments from 58 unique speakers, totaling \SI{3478}{\hour}, remain for further development.

\textbf{Italian} (\textsc{ita}) is a Romance language from the Italo-Dalmatian branch, with \SI{67}{M} total speakers, primarily in Italy. It is a non-tonal language with 23 consonants and 7 vowels. Compared to other Romance languages, Italian preserves much of the Vulgar Latin phonology. Italian is represented by the LaMIT Database \cite{DiBenedetto2022LaMITDatabaseRead}, licensed under CC BY 4.0.

\textbf{Catalan} (\textsc{cat}) is a Romance language from the Ibero-Romance branch, with \SI{9.3}{M} total speakers, primarily in Spain. It is a non-tonal language with 22 consonants, 7 vowels, and 4 diphthongs, inheriting the vowel system from Vulgar Latin. Represented by SLR~69~\cite{Kjartansson2020OpenSourceHighQuality}, Catalan features terminal devoicing, lenition, voicing assimilation, and vowel harmony.

\textbf{Galician} (\textsc{glg}) is also a Romance language in the Ibero-Romance branch, with \SI{3.4}{M} total speakers, primarily in Spain. Galician shares an ancestor with Portuguese (\textsc{por}), both originating from the Galician-Portuguese language, thus their significant mutual intelligibility. Galician features 21 consonants, 7 vowels, and 9 diphthongs \cite{Regueira1996Galician}, also inheriting the vowel system from Vulgar Latin. Like Catalan and some Portuguese dialects, Galician features vowel harmony. Galician is represented by SLR~67 \cite{Kjartansson2020OpenSourceHighQuality}.

\textbf{Spanish} (\textsc{spa}) is a Romance language from the Ibero-Romance branch, with \SI{559}{M} total speakers throughout the world, especially in Latin America, making it the second-most spoken language in terms of native speakers. Spanish is a non-tonal and syllable-timed language with 20 consonants, 5 vowels, and 5 diphthongs. Spanish phonology varies significantly across dialects and regions. Spanish data were drawn from six datasets recorded in Latin America \cite{Guevara-Rukoz2020CrowdsourcingLatinAmerican}, specifically, Argentina (SLR~61), Chile (SLR~71), Colombia (SLR~72), Peru (SLR~73), Puerto Rico (SLR~74), and Venezuela (SLR~75). SLR~61 includes a small amount of weather announcements in Peninsular Spanish. 

\textbf{Basque} (\textsc{eus}) is a language isolate with \SI{1}{M} total speakers, primarily in Spain. The only surviving member of the Paleo-European languages, Basque is a non-tonal language with 24 consonants, 5 vowels, and 6 diphthongs. It is represented by SLR~66 \cite{Kjartansson2020OpenSourceHighQuality}.

\textbf{Kazakh} (\textsc{kaz}) is Western Turkic language with \SI{16}{M} total speakers. Kazakh is a non-tonal language, with 18 consonants and 9 vowels. As with many Turkic languages, Kazakh exhibits vowel harmony. It is represented by the Kazakh Speech Dataset (SLR~140) \cite{Kadyrbek2023DevelopmentKazakhSpeech}, licensed under CC BY-SA 3.0.

\textbf{Ukrainian} (\textsc{ukr}) is an East Slavic language with \SI{39}{M} total speakers, primarily in Ukraine. It is a non-tonal language, with 32 consonants and 6 vowels. The data for Ukrainian is drawn from the Lada, Tetiana, Mykyta, and Oleksa text-to-speech (TTS) datasets in \cite{Smoliakov2023OpenSourceUkrainian}, all licensed under Apache 2.0.  

\textbf{Levantine Arabic} (\textsc{apc}) is a variety of Arabic (\textsc{ara} spoken in the Levant region, with \SI{54}{M} total speakers. While it shares similarities with the 28 Modern Standard Arabic (\textsc{arb}) consonants, vowels differ by dialects, usually with 5 short vowels, 5 long vowels, and 2 diphthongs \cite{Hitchcock2020BriefIntroductionSounds}. Data were drawn from the Arabic Speech Corpus \cite{Halabi2016ModernStandardArabic}, licensed under CC BY-SA 4.0, recorded by a single male speaker.


\textbf{Bengali} (\textsc{ben}) is an Eastern Indo-Aryan language from the Bengali-Assamese branch, with \SI{278}{M} total speakers, primarily in the Bengal region of the Indian subcontinent. As with most Indo-Aryan languages, it is non-tonal. It has 35 consonant \cite{Eberhard2024EthnologueLanguagesWorld}, 7 vowels all with nasalization \cite{Dasgupta2007Bangla}, and many diphthongs \cite{Masica1993IndoAryanLanguages}. Bengali data were drawn from SLR~37 \cite{Sodimana2018StepbyStepProcessBuilding}.

\textbf{Nepali} (\textsc{npi}) belongs to the Eastern Pahari branch, with \SI{32}{M} total speakers, primarily in Nepal. It is a non-tonal language, with 29 consonants, 6 oral vowels, 5 nasal vowels, and 10 diphthongs. Nepali data were drawn from SLR~43 \cite{Sodimana2018StepbyStepProcessBuilding}.

\textbf{Marathi} (\textsc{mar}) is a Southern Indo-Aryan language with \SI{99}{M} total speakers, primarily in the Maharashtra state in Western India. It is a non-tonal language, with more than 30 consonants and 12 vowels, including 2 diphthongs \cite{Pandharipande2007Marathi}. Marathi does not feature nasal vowels. Marathi data were drawn from SLR~64 \cite{He2020OpensourceMultispeakerSpeech}.

\textbf{Gujarati} (\textsc{guj}) is a Western Indo-Aryan language with \SI{63}{M} total speakers, primarily in the Gujarat state in Western India. It is a non-tonal language with 31 consonants, 8 vowels, and 2 diphthongs. Gujarati data were drawn from SLR~78 \cite{He2020OpensourceMultispeakerSpeech}.

\textbf{Malayalam} (\textsc{mal}) is a Southern Dravidian language with \SI{38}{M} total speakers, primarily in the Kerala state on the Southwestern coast of India. It is a non-tonal language, with 37 consonants, 11 vowels, and 2 diphthongs. As with other Dravidian languages, Malayalam features true subapical retroflexes. Malayalam is represented by SLR~63 \cite{He2020OpensourceMultispeakerSpeech}.

\textbf{Tamil} (\textsc{tam}) is a Southern Dravidian language with \SI{87}{M} total speakers, primarily in the Tamil Nadu state in Southern India. It is a non-tonal language with 18 consonants, 10 vowels, and 2 diphthongs. In addition to the true subapical retroflexes, Tamil features multiple rhotic consonants. Tamil is represented by SLR~65 \cite{He2020OpensourceMultispeakerSpeech}.

\textbf{Telugu} (\textsc{tel}) is a South-Central Dravidian language with \SI{96}{M} total speakers, primarily in the Andhra Pradesh and Telangana states in Southern India. A non-tonal language with 21 consonants and 11 vowels, it is represented by SLR~66 \cite{He2020OpensourceMultispeakerSpeech}.

\textbf{Kannada} (\textsc{kan}) is a Southern Dravidian language with \SI{59}{M} total speakers, primarily in the Karnataka state in Southwestern India. It is a non-tonal language with, 22 consonants, 20 vowels, and 2 diphthongs. It is represented by SLR~79 \cite{He2020OpensourceMultispeakerSpeech}.

\textbf{Mandarin Chinese} (\textsc{cmn}) is a Sino-Tibetan language with more than \SI{1.1}{B} total speakers, primarily in China. Mandarin Chinese is the most commonly spoken language in terms of native speakers. Chinese is a tonal language with 4 phonemic tones, 24 consonants, 8 vowels, and 6 diphthongs. The dataset for Mandarin Chinese is AISHELL-3 (SLR~93) \cite{Shi2021AISHELL3MultiSpeakerMandarin}, licensed under Apache 2.0.

\textbf{Japanese} (\textsc{jpn}) is a Japonic language with \SI{123}{M} total speakers, primarily in Japan. It is a non-tonal but pitch-accented language with 15 consonants, 5 vowels, and 3 diphthongs. Data for Japanese were drawn from emotional speech JVNV corpus \cite{Xin2024JVNVCorpusJapanese}, and the read speech subset of PJS Corpus \cite{Koguchi2020PJSPhonemebalancedJapanese}. Both datasets are licensed under CC BY-SA 4.0.

\textbf{Burmese} (\textsc{mya}) is a Sino-Tibetan language from the Tibeto-Burman branch with \SI{43}{M} total speakers, primarily in Myanmar. It is a tonal language with 3 primary tones, 31 consonants, 8 vowels, and 4 diphthongs. It is represented by SLR~80 \cite{Oo2020BurmeseSpeechCorpus}.

\textbf{Khmer} (\textsc{khm}) is an Austro-Asiatic language from the Mon-Khmer family with \SI{18}{M} total speakers, primarily in Cambodia. Unlike most continental Southeast Asian languages, it is a non-tonal language. Khmer has 44 consonants and 24 vowels \cite{SokVathanak2023BasicKhmer}. Khmer is represented by SLR~42 \cite{Sodimana2018StepbyStepProcessBuilding}.

\textbf{Malay} (\textsc{zlm}) is an Austronesian language from the Malayo-Polynesian branch with \SI{19}{M} total speakers, primarily in Malaysia. It is part of the Malay macrolanguage (\textsc{msa}) used throughout maritime Southeast Asia that includes Standard Malay (\textsc{zsm}) and Indonesian (\textsc{ind}). Like most Austronesian languages, Malay is non-tonal, with 19 consonants and 6 vowels \cite{Abdullah1972MorphologyMalay}. Data for Malay were drawn from datasets released via Malaya Speech \cite{Husein2024MalayaSpeech}, predominantly recorded by one male speaker, all licensed under CC BY 4.0.

\textbf{Javanese} (\textsc{jav}) is an Austronesian language from the Malayo-Polynesian branch with \SI{68}{M} total speakers, primarily in the Java and Sumatra islands of Indonesia. Javanese is a non-tonal language with 21 consonants and 8 vowels. It is represented by SLR~41 \cite{Sodimana2018StepbyStepProcessBuilding}.

\textbf{Sundanese} (\textsc{sun}) is an Austronesian language from the Malayo-Polynesian branch with SI{37}{M} total speakers, primarily in the Western part of the Java island in Indonesia. Sundanese is a non-tonal language with 18 consonants and 7 vowels. Sundanese is represented by SLR~44 \cite{Sodimana2018StepbyStepProcessBuilding}.

\textbf{Afrikaans} (\textsc{afr}) is a West Germanic language with 18M speakers, primarily in South Africa. It is highly related to Dutch (\textsc{nld}) due to the Dutch colonization of South Africa. It is a non-tonal language with 20 consonants, 16 vowels, and 9 diphthongs. Afrikaans is represented by SLR~32 \cite{Niekerk2017RapidDevelopmentTTS}.

\textbf{Xhosa} (\textsc{xho}) is a Bantu language from the Nguni branch, with \SI{19}{M} total speakers, primarily in South Africa. It is a tonal language with 2 phonemic tones (high and low) and 10 vowels. Xhosa notably has a complex system of consonants, with multiple ejective stops and click consonants \cite{Jessen2002VoiceQualityDifferences}. It is also represented by SLR~32 \cite{Niekerk2017RapidDevelopmentTTS}.

\textbf{Sesotho} (\textsc{sot}) is a Bantu language from the Sotho-Tswana branch, with \SI{13.5}{M} total speakers, primarily in South Africa and Lesotho. It is a tonal language with 2 tones, more than 30 consonants including 3 clicks, and 9 vowels \cite{Paroz1946ElementsSouthernSotho}. It is also represented by SLR~32 \cite{Niekerk2017RapidDevelopmentTTS}.

\textbf{Setswana} (\textsc{tsn}) also belongs to the Sotho-Tswana branch, with \SI{14}{M} user, primarily in South Africa and Botswana. It is a tonal language with two tones, 29 consonants, and 7 vowels \cite{Bennett2016SetswanaSouthAfrican}. It is also represented by SLR~32 \cite{Niekerk2017RapidDevelopmentTTS}.

\textbf{Yoruba} (\textsc{yor}) is an Edekiri language with \SI{47}{M} total speakers, primarily in Nigeria. A tonal language with 3 tones, 17 consonants, 7 oral vowels, and 4 nasal vowels, Yoruba features vowel harmony. It is represented by SLR~86 \cite{Gutkin2020DevelopingOpenSourceCorpus}.

\FloatBarrier
\section{Dataset Generation} \label{sec:dataset-gen}

The process for generating the stem track is broadly similar to that of DnR v2. The precise processes are detailed in \Cref{algo:stem} and \Cref{algo:time}. The process at the high level is as follows. 

For each stem, the track-level loudness $L_{\text{track}}$ and number of sound events $N_{\text{event}}$ were randomly drawn. For each sound event, up to 10 event candidates were randomly drawn from the pool of raw sound events and checked against timing constraints. The first raw event that satisfies the timing constraints is taken. The timing parameters (start time within track $t_i$, start time within event $\eta_i$, and event duration $U_i$) were then set or randomized accordingly. If none of the candidates satisfy the constraints, that event is skipped. The event-level loudness $L_{\text{event}, i}$ was then drawn from a distribution centered at the track-level loudness. The raw sound event was segmented and loudness-normalized accordingly, before being added to the stem. Once all events had been added, the stem was loudness normalized again to the track-level loudness. Finally, mastering is applied to meet target loudness and peak specifications as detailed in \Cref{sec:dataset-gen/mastering}.
All mentions of ``loudness'' in this work refer to the integrated loudness as defined in ITU-R BS.1770 \cite{InternationalTelecommunicationUnion2015ITURBS17704}. 

We will now discuss the details and motivation behind each step and any changes from version 2. Stem-dependent parameters are provided in \Cref{tab:genparam}. 

\begin{table}[t]
    \centering
    \caption{Stem-dependent parameters}\vspace{-5pt}
    \begin{tabular}{l@{ }l*{4}{S[table-format=-2.2]}}
    \toprule
    \multicolumn{2}{l}{Parameter} & {DX} & {MX} & {FGFX} & {BGFX} \\
    \midrule
    $\lambda_{\text{event}}$ & & 12.0 & 7.0 & 12.0 & 24.0 \\
    \midrule
    $\Delta \mu_{\text{track}}$ & (LKFS)
        & 0.0 & -5.0 & -5.0 & -13.0 \\
    $\sigma_{\text{track}}$ & (LKFS)
        & 4.0 & 6.0 & 6.0 & 6.0 \\
    $\sigma_{\text{event}}$ & (LKFS)
        & 6.0 & 10.0 & 10.0 & 10.0 \\
    \midrule 
    $T_{\text{min}}$ & (s) & 0.0  & 0.0  & 0.5 & 1.0 \\
    $\beta_{\text{min}}$ & & 1.0 & 0.3 & 0.3 & 0.3\\
    $\gamma$ & & 0.75 & 1.0 & 0.5 & 0.0 \\
     \bottomrule
    \end{tabular}
    
    \label{tab:genparam}
\end{table}

\begin{algorithm}[t]
    \caption{Stem Generation}
    \label{algo:stem}

    \footnotesize

    Initialize buffer $\mathbf{s}$ to zeros of duration $D_{\text{track}}$.\\
    Initialize track cursor $t_{\text{cur}} \gets \SI{0}{\second}.$ \\
    \BlankLine
    
    Sample track loudness $L_{\text{track}} \sim \mathcal{N}(\mu_\text{{ref.}} + \Delta\mu_{\text{track}}, \sigma^2_{\text{track}})$.\\
    Sample number of events $N_{\text{event}} \sim \mathcal{ZTP}(\lambda_{\text{event}})$.
    \BlankLine
   
    \For{$i \coloneqq 1$ to $N_{\text{event}}$}{
        \For{up to 10 trials}{
            Sample raw sound event $\mathbf{z}_i \sim \mathfrak{Z} $.\\
            Run \Cref{algo:time} on $\mathbf{z}_i$ to check and get timing parameters.\\
            \If{\Cref{algo:time} succeeds}{
                Set $t_i$, $\eta_i$, and $U_i$ accordingly.\\

                \BlankLine

                Sample event loudness $L_{\text{event}, i} \sim \mathcal{N}(L_{\text{track}}, \sigma_{\text{event}}^2)$. \\
                Segment $\tilde{\mathbf{z}}_i \gets \mathbf{z}_i(\eta_i \colon \eta_i + U_i)$\\
                Normalize loudness of $\tilde{\mathbf{z}}_i$ to $L_{\text{event}, i}$.\\
        
                \BlankLine
                
                Set $\mathbf{s}(t_i \colon t_i + U_i) \gets \mathbf{s}(t_i \colon t_i + U_i) + \tilde{\mathbf{z}}_i$.\\
                \BlankLine
                
                Sample $\Delta t_{\text{cur}} \sim \mathcal{U}(\gamma U_i, U_i)$.\\
                Set $t_{\text{cur}} \gets t_{\text{cur}} + \Delta t_{\text{cur}}$.
                \BlankLine
                
                \textbf{break}
            }
        }


    }
    \BlankLine
    
    Normalize loudness of $\mathbf{s}$ to $L_{\text{track}}$.
   
\end{algorithm}

\begin{algorithm}[t]
    \caption{Check and get timing parameters}
    \label{algo:time}

    \footnotesize

    \textbf{Input:} Raw event $\mathbf{z}_i$, Cursor $t_{\text{cur}}$.
    \BlankLine

    Set $T_i \gets$ duration of $\mathbf{z}_i$, in seconds.\\

    Set $S_{\text{min}} \gets \max\{T_{\text{min}}, \beta T_i\}$.\\
    Set $t_{\text{max}} \gets D_{\text{track}}  - S_{\text{min}}$.
    
    \BlankLine
    \lIf{$D_{\text{track}} - t_{\text{cur}} < S_{\text{min}}$}{
        \Return \textsc{fail}.
    }
    \BlankLine

    \If{$\beta_{\text{min}} < 1$}{
        Clamp $t_{\text{max}} \gets \min\{t_{\text{max}}, D_{\text{track}} - \delta\}$.\\
        \lIf{$t_{\text{max}} < t_{\text{cur}}$}{
            \Return \textsc{fail}.
        }
    }

    \BlankLine
    
    Sample start time in track $t_i \sim \mathcal{SN}(t_{\text{cur}}, \rho^2_{\text{start}}, \alpha)$.\\
    \BlankLine

    \uIf{$\beta_{\text{min}} = 1$ \textbf{or} $T_i \le S_{\text{min}}$}{
        Clamp $t_i$ to range $[t_{\text{cur}}, t_{\text{max}}]$.\\
        Set event duration $U_i \gets T_i$.\\
    }\Else{
        Set $S_{\text{max}} \gets \min\{T_i, D_{\text{track}} - t_i\}$\\
        \lIf{$S_{\text{max}} < 0$}{\Return \textsc{fail}.}
        \BlankLine
        Clamp $t_i \gets \max\{t_i, \SI{0}{\second}\}$. \\
        Sample event duration $U_i \sim \mathcal{TN}(\upsilon T_i, (\rho T_i)^2, S_{\text{min}}, S_{\text{max}})$.
    }
        
    \BlankLine
    
    \uIf{MX}{
        Sample event start time offset $\eta_i \sim \mathcal{U}(\SI{0}{\second}, T_i - U_i)$.
    }\Else{
        Set event start time offset $\eta_i = \SI{0}{\second}$.
    }

    \Return $t_i$, $\eta_i$, $U_i$
   
\end{algorithm}

\begin{figure}[t]
    \centering
    \includegraphics[width=\columnwidth]{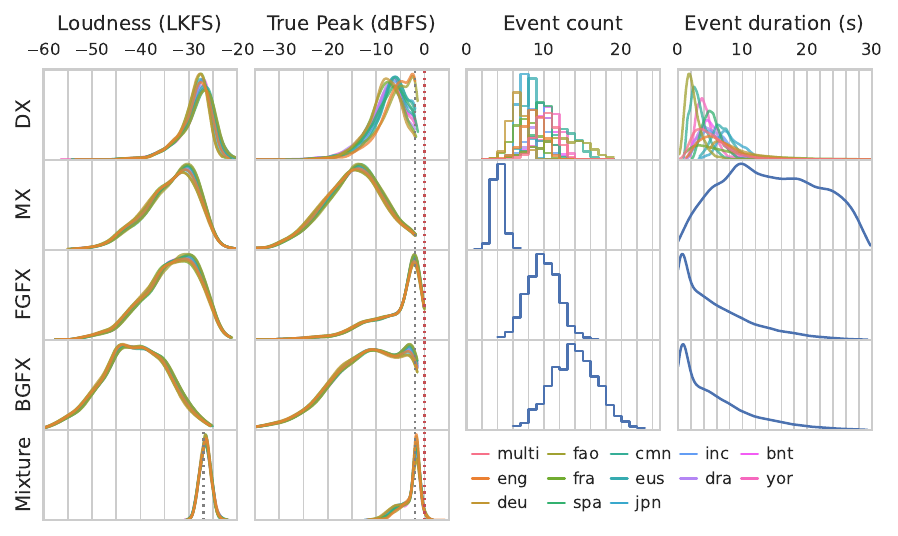}\vspace{-8pt}
    \caption{Test set distribution of the post-mastering loudness, true peak, event counts, and event durations for each stem. Each colored line represents a variant. The dotted line in the mixture loudness plot indicates the \SI{-27}{LKFS} level. The dotted lines in the peak plots indicate \SI{-2}{dBFS} and \SI{0}{dBFS} levels.}
    \label{fig:distr}
\end{figure}

\subsection{Loudness distribution} \label{sec:dataset-gen/loudness}

In DnR v3, the track-level loudness $L_{\text{track}}$ is drawn from $\mathcal{N}(\mu_\text{{ref}} + \Delta\mu_{\text{track}}, \sigma^2_{\text{track}})$, 
where $\mu_\text{{ref}} = \SI{-27}{LKFS}$ is an arbitrary reference level,
$\Delta\mu_{\text{track}}$ is a stem-dependent mean loudness level,
and $\sigma_{\text{track}}$ is a stem-dependent track-level loudness standard deviation (SD). 
Unlike DnR v2, $\mu_\text{{ref}}$ does not affect peak limiting during mastering. The event-level loudness $L_{\text{event}}$ was drawn from $\mathcal{N}(L_{\text{track}}, \sigma^2_{\text{event}})$, where 
$\sigma_{\text{event}}$ is a stem-dependent event-level loudness SD. Note that once the entire stem is generated, the entire track is normalized back to $L_{\text{track}}$.

For simplicity, we set $\Delta\mu_{\text{track}} = \SI{0}{LKFS}$ for the DX stem, $\Delta\mu_{\text{track}} = \SI{-13.0}{LKFS}$ for the BGFX stem, and $\Delta\mu_{\text{track}} = \SI{-5}{LKFS}$ for the music and FGFX stem. The level for music and foreground stems relative to the DX stem are approximately based on the observed distributions both in Netflix's own content distribution and in CDXDB23 \cite[Tab\@. 6]{Uhlich2023SoundDemixingChallenge} which was derived from Sony content. Setting the BGFX stem at \SI{8}{LKFS} under the FGFX is based on DnR v2. Event loudness normalization is performed using pyloudnorm \cite{Steinmetz2021PyloudnormSimpleFlexible}, which implements ITU-R BS.1770-4 \cite{InternationalTelecommunicationUnion2015ITURBS17704}.

Two major changes from version 2 are in the significantly increased spread of the loudness distribution and the underlying distribution itself. In DnR v2, $L_{\text{track}}$ was drawn from a uniform distribution with bounds $\pm \SI{2}{LKFS}$ from the mean, while $L_{\text{event}}$ was drawn uniformly with bounds $\pm \SI{1}{LKFS}$ from $L_{\text{track}}$. The resulting triangle distribution can be seen to be much narrower than the distribution of the loudness in CDXDB23 \cite[Fig\@. 6]{Uhlich2023SoundDemixingChallenge}. To better approximate the loudness distribution, we switched to the normal distribution in version 3. The fairly large event-level SDs were set so as to also help mimic the effects of panning on the loudness, hopefully allowing the models trained on it to better function in a pseudo-stereo mode.

\subsection{Event density} \label{sec:dataset-gen/events}

As with DnR v2, the number of sound events $N-{\text{event}}$ within each track is drawn from a zero-truncated Poisson distribution $\mathcal{ZTP}(\lambda_{\text{event}})$, where $\lambda_{\text{event}}$ is a stem-dependent parameter for the mean number of events. The values of $\lambda_{\text{event}}$ for MX and FGFX are the same as those of DnR v2. The value of $\lambda_{\text{event}}$ for BGFX is set very high, so that, in tandem with the low relative loudness, the stem provides a dense but unsalient backdrop to the mixture. The value of $\lambda_{\text{event}}$ for DX is set slightly higher than that of DnR v2 to account for the relatively shorter average length of raw sound events in some of the contributing dialogue datasets. In practice, it can be observed that the values of $\lambda_{\text{event}}$ are higher than the actual mean number of events that can fit into each track given the timing constraints. Nonetheless, the distributions of the number of events generally follow Poisson-like shapes. 

\subsection{Sound event sampling}
For each event, a random raw sound event candidate $\mathbf{z}_i$ is drawn randomly from the pool of possible sound events $\mathfrak{Z}$, where $i$ is the event index. Sampling for dialogue events in monolingual variants, all music events, and all effects events were done uniformly with replacement. For the Indo-Aryan (\textsc{inc}), Dravidian (\textsc{dra}), and Bantu (\textsc{bnt}) variants, a language is drawn with a probability defined by the ratio of the total number of speakers of the language relative to the total number of speakers of all represented languages in the respective variant. For the \textsc{multi} variant, languages from the West Germanic and Romance families were drawn according to the total number of speakers of each language. All other languages were drawn with a probability proportional to the total number of speakers of the language family. Within each language, each sample is drawn with an equal probability.

\subsection{Timing Parameters} \label{sec:dataset-gen/timing}

For each sound event candidate, the process detailed in \Cref{algo:time} is applied to check whether the event can be placed into the stem, and if so, to compute the timing parameters, namely, the start time relative to the track $t_i$, the start time relative to the raw sound event $\eta_i$, and the event duration $U_i$. This process is very similar to that of DnR v2, except that a non-music sound event was allowed to overlap with the previous event by a duration of up to $(1-\gamma)U_{i-1}$, where $\gamma$ is stem-dependent. At the start of the process, a track cursor $t_{\text{cur}}$ is set to \SI{0}{\second}. After each event is added, $t_{\text{cur}}$ is advanced by a random amount drawn from $\mathcal{U}(\gamma U_i, U_i)$. 

The first constraint is that the remaining duration in a track is at least $U_{\text{min}} = \max\{T_{\text{min}}, \beta_{\text{min}} T_i\}$, where  $T_{\text{min}}$ is the absolute minimum duration, $\beta_{\text{min}}$ is the minimum proportional duration, and $T_i$ is the duration of $\mathbf{z}_i$. For the DX stem where $\beta_{\text{min}} = 1$, this ensures that the entire raw sound event must fit in the track to prevent a word from being cut off mid-enunciation. For the rest of the stems where $\beta_{\text{min}} < 1$, the latest possible start time $t_{\text{max}} = \min\{D_{\text{track}} - U_{\text{min}}, D_{\text{track}} - \delta\}$ must be later than $t_{\text{cur}}$, which approximately defines the earliest possible start time. The parameter $\delta = \SI{2.0}{\second}$ follows from v2, and defines the minimum time before the end of the track that a non-dialogue event could start. 

Once the first two constraints are satisfied. Similar to DnR v2, The start time relative to the track $t_i$ is sampled from a skew-normal distribution $\mathcal{SN}(t_{\text{cur}}, \rho^2_{\text{start}}, \alpha)$ where $\rho_{\text{start}} = \SI{2.0}{\second}$ defines the spread of the distribution and $\alpha = 5.0$ defines the skew parameter of the distribution. Since the skew-normal distribution is unbounded, $t_i$ can take a value less than $t_{\text{cur}}$ or larger than $D_{\text{track}}$, although with low probabilities. For the DX stem or events where $T_{\text{min}} < U_{\text{min}}$, $t_i$ is clamped to the range $[t_{\text{cur}}, t_{\text{max}}]$ and the entire raw sound event is added to the stem, i.e. $U_i = T_i$. For other cases, if $t_i$ is larger than $D_{\text{track}}$, we simply draw a new candidate. For the cases where $t_i < D_{\text{track}}$, we clamp $t_i$ to ensure $t_i \ge \SI{0}{\second}$. Following DnR v2, the event duration $U_i$ is then drawn from a truncated normal distribution $\mathcal{TN}(\upsilon_{\text{dur}} T_i, (\rho_{\text{dur}} T_i)^2, U_{\text{min}}, U_{\text{max}})$ where $\upsilon_{\text{dur}} T_i$ is the location parameter, $\rho_{\text{dur}} T_i$ is the spread parameter, $U_{\text{min}}$ is the lower bound, and $U_{\text{max}}$ is the upper bound, with $\upsilon_{\text{dur}} = 0.5$ and $\rho_{\text{dur}} = 0.1$. For the MX stem with $U_i < T_i$, the start time offset $\eta_i$ is randomized from a uniform distribution $\mathcal{U}(\SI{0}{\second}, T_i - U_i)$. All other stems start from the beginning of the raw sound event, i.e. $\eta_i = \SI{0}{\second}$.

\subsection{Mastering} \label{sec:dataset-gen/mastering}

Following Netflix's specifications of mastering to \SI[separate-uncertainty=true]{-27(1)}{LKFS}, we draw the target loudness level from $\mathcal{N}(\mu_{\text{mix}}, \sigma^2_{\text{mix}})$ with $\mu_{\text{mix}} = \SI{-27}{LKFS}$ and $\sigma_{\text{mix}} = \SI{1}{LKFS}$. The choice of drawing the loudness randomly instead of targeting \SI{-27}{LKFS} directly is to reflect the occasional non-compliance with the delivery specifications seen in practice. We follow the true peak limit specification for Netflix and Hulu, which is \SI{-2}{dBFS}.

In practice, loudness normalization and peak limiting are usually applied on the mix bus. However, these processes are nonlinear and often time-variant. In order to keep the mixture as a linear sum of the stems, we apply loudness normalization and peak limiting to each stem individually. 

In DnR v2, peak limiting was done using naive (sample) peak computation followed by attenuation of the entire stem. This, however, is not a common practice and can severely distort the relative loudness distribution between stems. In DnR v3, we first compute an intermediate mixture based on the generated stems. Using this intermediate mixture, we compute the necessary gain adjustment needed to get the mixture to the target loudness. Using this value, each stem is then linearly scaled to meet their respective absolute target loudness using pyloudnorm. We then use the loudnorm filter of FFmpeg to compute the necessary filter parameters in the first pass, and to jointly apply EBU R 128 loudness normalization and peak limiting \cite{EuropeanBroadcastingUnion2023EBU1282023Loudness} in the second pass, obtaining the final stems. Dynamic normalization is allowed if the first pass of loudnorm indicates that linear time-invariant normalization cannot achieve the required specifications. The final mixture is then computed by linearly combining the stems.  Within any split of any variant, at most \SI{2.7}{\percent} and \SI{2.0}{\percent} of the mixtures are clipped in terms of true peak and sample peak, respectively.

\section{Experimental Setup} \label{sec:expt}

\subsection{Model}  \label{sec:expt/model}
The architecture of the model used in this work is identical to the 64-band music variant of Bandit in \cite{Watcharasupat2023GeneralizedBandsplitNeural}. Note that the model in \cite{Watcharasupat2023GeneralizedBandsplitNeural} was trained on data sampled at \SI{44.1}{\kilo\hertz} while the models in this work were trained on data sampled at \SI{48}{\kilo\hertz}. As a result, while the band definition in terms of the discrete frequency domain indices does not change, the actual band edges in terms of the absolute frequency in Hz are different.

\subsection{Training and Testing}  \label{sec:expt/trtt}

All models in this work were trained using the following setup, which is slightly different from \cite{Watcharasupat2023GeneralizedBandsplitNeural}. Each model was trained for 200 epochs with a batch size of 8 per GPU, using an AdamW optimizer \cite{Loshchilov2019DecoupledWeightDecay} with an initial learning rate of \num{e-3} and a decay factor of \num{0.99} per epoch. Each epoch consisted of 2048 batches per GPU. For each training clip, a random chunk of \SI{8.0}{\second} was drawn for each stem independently of other stems. The training mixture chunks were then recomputed from the random stem chunks. Testing was done on the entire track, using overlap-add in the same way as in \cite{Watcharasupat2023GeneralizedBandsplitNeural}, with a chunk size of \SI{8.0}{\second} and a hop size of \SI{1.0}{\second}. 

Each model was trained using 8 NVIDIA A100 Tensor Core GPUs (40 GB each) on an Amazon EC2 {p4d.24xlarge} instance. Testing and inference were done using NVIDIA T4 Tensor Core GPUs on an Amazon EC2 {g4dn.metal} instance. An updated implementation of Bandit in this work employs significant gradient checkpointing to reduce the memory footprint, thus significantly increasing the chunk size and batch size per GPU memory, compared to \cite{Watcharasupat2023GeneralizedBandsplitNeural}.\footnote{We attempted to use mixed precision through multiple approaches, including only in parts of the model. However, none of the attempted approaches were sufficiently stable during training, thus were not employed in this work.}

\section{Results and Discussion}  \label{sec:results}

For brevity, we report only the median track-level SNRs in the paper. Mean SNR, mean SI-SNR, and median SI-SNR follow a similar trend to median SNR. Raw clip-wise SNR and SI-SNR are available in the repository. The median SNR results are provided in \Cref{tab:perf}.  

\begin{table*}[t]
    \centering
    \setlength{\tabcolsep}{1.5pt}
    \renewcommand{\arraystretch}{0.7}
    \caption{Performance of Bandit Models on Different Variants of DnR v3}\vspace{-5pt}
    \begin{tabularx}{\linewidth}{>{\scshape}X *{21}{S[table-format=-2.1]}}
    \toprule
    & \multicolumn{7}{c}{Dialogue SNR (dB), by Train Variant}
    & \multicolumn{7}{c}{Music SNR (dB), by Train Variant}
    & \multicolumn{7}{c}{Effects SNR (dB), by Train Variant}\\
    \cmidrule(lr){2-8}
    \cmidrule(lr){9-15}
    \cmidrule(lr){16-22}
    \multicolumn{1}{l}{Test Var\@.}
    & \multicolumn{1}{r}{\textsc{multi}}
    & \multicolumn{1}{r}{\textsc{eng}} 
    & \multicolumn{1}{r}{\textsc{deu}} 
    & \multicolumn{1}{r}{\textsc{fao}} 
    & \multicolumn{1}{r}{\textsc{fra}} 
    & \multicolumn{1}{r}{\textsc{spa}} 
    & \multicolumn{1}{r}{\textsc{cmn}}
    & \multicolumn{1}{r}{\textsc{multi}}
    & \multicolumn{1}{r}{\textsc{eng}} 
    & \multicolumn{1}{r}{\textsc{deu}} 
    & \multicolumn{1}{r}{\textsc{fao}} 
    & \multicolumn{1}{r}{\textsc{fra}} 
    & \multicolumn{1}{r}{\textsc{spa}} 
    & \multicolumn{1}{r}{\textsc{cmn}}
    & \multicolumn{1}{r}{\textsc{multi}}
    & \multicolumn{1}{r}{\textsc{eng}} 
    & \multicolumn{1}{r}{\textsc{deu}} 
    & \multicolumn{1}{r}{\textsc{fao}} 
    & \multicolumn{1}{r}{\textsc{fra}} 
    & \multicolumn{1}{r}{\textsc{spa}} 
    & \multicolumn{1}{r}{\textsc{cmn}} \\
    \midrule
multi & \bfseries 15.8 & 15.1 & 14.9 & 2.5 & 14.9 & 7.4 & 3.2 & 10.3 & 10.3 & \bfseries 10.4 & 8.6 & 10.3 & 9.0 & 8.7 & \bfseries 10.0 & 9.5 & 9.6 & -0.3 & 9.5 & 4.7 & 1.1 \\
    \midrule
eng & 15.3 & \bfseries 15.6 & 14.9 & 2.5 & 14.4 & 8.9 & 1.2 & \bfseries 10.4 & \bfseries 10.4 & \bfseries 10.4 & 8.3 & \bfseries 10.4 & 8.7 & 8.0 & \bfseries 9.9 & \bfseries 9.9 & 9.8 & -0.1 & 9.3 & 6.7 & -0.9 \\
deu & 15.3 & \bfseries 15.4 & 15.2 & 0.9 & 15.3 & 8.8 & 0.5 & 9.8 & \bfseries 10.0 & \bfseries 10.0 & 7.1 & \bfseries 10.0 & 7.6 & 6.3 & 9.5 & 9.4 & \bfseries 9.6 & -3.2 & 9.5 & 5.9 & -3.0 \\
fao & 14.0 & 12.7 & 12.8 & \bfseries 15.1 & 13.0 & 6.5 & 2.9 & 10.4 & 10.4 & \bfseries 10.5 & 10.3 & 10.4 & 9.9 & 9.8 & 9.3 & 8.3 & 8.6 & \bfseries 10.2 & 8.7 & 4.4 & 1.7 \\
fra & 15.3 & 14.0 & 14.3 & 0.2 & \bfseries 15.7 & 3.0 & 0.6 & 9.7 & 9.6 & 9.7 & 6.2 & \bfseries 10.0 & 5.7 & 5.6 & 9.5 & 8.7 & 9.0 & -3.7 & \bfseries 9.8 & 0.2 & -2.7 \\
spa & \bfseries 16.2 & 15.9 & 15.5 & 4.6 & 15.3 & \bfseries 16.2 & 3.4 & 10.8 & 10.8 & \bfseries 10.9 & 9.9 & 10.8 & 10.8 & 9.9 & \bfseries 10.6 & 10.3 & 10.4 & 2.9 & 10.3 & 10.5 & 2.6 \\
cmn & \bfseries 15.5 & 13.4 & 13.6 & 0.4 & 14.0 & 2.2 & 15.4 & \bfseries 10.6 & 10.4 & 10.5 & 9.0 & 10.5 & 9.3 & \bfseries 10.6 & \bfseries 10.3 & 9.2 & 9.6 & -1.4 & 9.6 & 0.5 & 10.2 \\
\midrule
eus & \bfseries 15.0 & 14.2 & 13.9 & 2.3 & 14.0 & 13.8 & 6.3 & 10.3 & 10.3 & \bfseries 10.4 & 9.3 & \bfseries 10.4 & 9.8 & 9.7 & \bfseries 10.2 & 9.7 & 9.8 & 0.3 & 9.7 & 9.5 & 4.7 \\
jpn & \bfseries 14.9 & 9.0 & 8.9 & 2.5 & 9.7 & 3.9 & 0.6 & \bfseries 10.2 & 10.0 & 10.1 & 8.6 & 10.1 & 9.6 & 9.2 & \bfseries 9.5 & 5.5 & 5.9 & 0.1 & 6.3 & 1.2 & -2.2 \\
inc & 16.8 & \bfseries 16.9 & 16.5 & 2.6 & 16.2 & 15.6 & 2.1 & 10.7 & 10.8 & \bfseries 10.9 & 9.3 & 10.8 & 10.5 & 9.6 & 10.5 & 10.4 & \bfseries 10.6 & 0.2 & 10.4 & 10.0 & -0.0 \\
dra & 16.9 & \bfseries 17.0 & 16.6 & 6.2 & 16.2 & 16.4 & 3.0 & 10.8 & 10.8 & \bfseries 10.9 & 9.8 & 10.8 & 10.6 & 9.7 & 10.6 & 10.5 & \bfseries 10.7 & 3.7 & 10.4 & 10.3 & 1.1 \\
bnt & \bfseries 16.3 & 14.3 & 14.1 & 12.1 & 14.3 & 12.3 & 5.8 & \bfseries 10.6 & 10.3 & 10.5 & 10.0 & 10.4 & 10.0 & 9.9 & \bfseries 10.3 & 9.0 & 9.1 & 8.1 & 9.1 & 8.1 & 3.8 \\
yor & \bfseries 16.2 & 14.4 & 14.6 & 7.1 & 15.6 & 9.3 & 9.7 & \bfseries 10.6 & 10.3 & 10.5 & 9.8 & \bfseries 10.6 & 9.8 & 10.1 & \bfseries 10.6 & 9.9 & 10.1 & 5.1 & 10.5 & 6.7 & 7.6 \\
    \bottomrule
    \end{tabularx}
    \label{tab:perf}
    
    Bold numbers indicate the best-performing model(s) by stem and testing variant.
\end{table*}



\subsection{Monolingually Trained Models}  \label{sec:results/mono}

To first demonstrate the issues of using monolingually trained models, we trained six models using monolingual variants of DnR v3 in \textsc{eng}, \textsc{deu}, \textsc{fao}, \textsc{fra}, \textsc{spa}, and \textsc{cmn}. These languages were chosen as training variants as they consist of more than 30 hours of total data. We then test each model exhaustively on the test set of the aforementioned six monolingual variants, additional three monolingual test sets (\textsc{eus}, \textsc{jpn}, and \textsc{yor}), and three multilingual test sets grouped by language family (\textsc{inc}, \textsc{dra}, and \textsc{bnt}). 

The in-language DX performance exceeds \SI{15}{dB} SNR in all six languages. Similarly, in-language MX and FX SNRs all exceed \SI{10}{dB} and \SI{9.5}{dB}, respectively. For the DX and FX stems, this is a similar result to the model trained on English-only DnR v2 as reported in \cite{Watcharasupat2023GeneralizedBandsplitNeural}. The increase in MX SNR is likely due to the removal of vocals from DnR v2. 
For cross-language evaluation, however, the \textsc{fao}- and \textsc{cmn}-trained variants struggled to generalize to other languages, performing at under \SI{8}{dB} SNR for all but one test set each. The \textsc{eng}-, \textsc{deu}-, \textsc{fra}-trained models appear to generalize significantly better, struggling slightly with \textsc{fao} and more so with \textsc{jpn} test sets. The \textsc{spa}-trained model appears to perform well on \textsc{eus}, \textsc{inc}, \textsc{dra}, and \textsc{bnt} test sets, but struggled with the rest. 

It is important to note, however, that due to the various sources each language drew its data from, the performance variations cannot be solely explained by linguistic differences. Notably, \textsc{eng}, \textsc{deu}, \textsc{fra} variants drew their data from crowdsourced audiobook websites. As a result, their data contains significantly more diverse sets of acoustic environments and recording setups. Raw source data for \textsc{spa}, \textsc{eus}, \textsc{inc}, \textsc{dra}, \textsc{bnt}, and \textsc{yor} were all collected as a part of a Google initiative, thus likely share similar recording hardware and portable studio setups, albeit in different geographical locations \cite[Sect\@. 2.2]{Butryna2020GoogleCrowdsourcedSpeech}. AISHELL-3, the source dataset for \textsc{cmn}, appears to have been recorded in a very uniform manner, likely in a single acoustic environment \cite[Sect\@. 2]{Shi2021AISHELL3MultiSpeakerMandarin}. Similarly, Faroese BLARK \cite{Simonsen2022CreatingBasicLanguage}, the source dataset for \textsc{fao}, was always recorded using a TASCAM DR-40, although in a few different acoustic environments \cite{Debess2022SoundRecordingsProject}. Having a very uniform acoustic environment and recording setup is usually beneficial for TTS, the original purpose of these datasets. Our somewhat unintended use case in source separation might indicate that a diversity of acoustic environment and recording setup is an additional important factor for CASS model generalizability.

\subsection{Multilingually Trained Model} \label{sec:results/multi}

Next, we investigate the performance of the multilingually trained model, which has training materials from all languages listed in \Cref{tab:lang}. Note that the number of hours of DX content seen by the model is identical (\SI{100}{\hour}) to monolingual models; the \SI{100}{\hour} is now shared between 32 languages instead of being dedicated to one language.

The multilingual model generally performed well across all languages, with upwards of \SI{14}{dB} SNR on DX, \SI{9.7}{dB} SNR on MX, and \SI{9.3}{dB} on FX. For the languages with dedicated training sets, the multilingual model performs either better or within \SI{1.1}{dB} of their respective monolingual model on DX, \SI{0.3}{dB} on MX, and \SI{0.9}{dB} on FX. For the test sets without corresponding training sets, the multilingual model either performs the best or within \SI{0.2}{dB} of the best-performing monolingual variants (either \textsc{eng} or \textsc{deu}) across all stems. Moreover, the multilingual model significantly outperforms all monolingual variants on \textsc{jpn} in the DX stem by at least \SI{5}{dB} and in the FX stem by at least \SI{3.5}{dB}, despite very limited \textsc{jpn} training materials. In general, this demonstrates that the multilingual model can consistently perform well across multiple languages, and is usually on par or better with dedicated monolingual models, even with comparatively less language-specific data to the language. This means that it is possible to only use one model to perform CASS in many languages without needing to train many dedicated models. This would particularly benefit low-resource languages with insufficient data to create a training set. Admittedly, the \textsc{eng} and \textsc{deu} also perform similarly or slightly better than the multilingual model on some test sets. However, these two monolingual models are less consistent than multilingual models in terms of their performance across languages.

\section{Conclusion} \label{sec:concl}

In this work, we present version 3 of the Divide and Remaster dataset, addressing some of the areas of improvement that have been identified in DnR v2. Specifically, DnR v3 tackled issues with regard to vocals and vocalization in music and effects stem, loudness distribution, mastering, and audio formats. Moreover, while DnR v2 is an English-only dataset, the dialogue stem in DnR v3 drew its content from 32 languages. Benchmarking experiments using Bandit demonstrated that a multilingually trained model can perform on par or close to dedicated monolingual models, enabling CASS on contents in languages with low data availability. 

Despite our efforts, however, CASS remains in an early stage of research, with significant areas for future work both in the dataset curation and in the model development that are out of scope for this paper. Specifically, \begin{inparaenum}
    \item a significant number of languages or even entire language families remain missing from DnR v3;
    \item emotional diversity within the dialogue stem remains fairly limited in most languages;
    \item spatialization, equalization, reverberation, and several other production aspects of cinematic contents remain unaddressed; and 
    \item nuances around the issues of non-speech human vocalized sounds continue to evade the strict three-stem setup.
\end{inparaenum}
The authors are actively working to address these issues and welcome any effort to contribute to future iterations of the dataset.



\section*{Acknowledgment}

The authors thank the following people and teams who have assisted with the project in various ways: 
Darius P{\'e}termann;
Gordon Wichern and the Mitsubishi Electric Research Laboratories (MERL) Speech \& Audio team;
Pablo Delgado and the Netflix Machine Learning Platform team;
Scott Kramer and the Netflix Sound Technology team; 
Phillip A. Williams, William Wolcott, and the Netflix Audio Algorithms team. 
 
Part of this work was done while K. N. Watcharasupat was separately supported by the IEEE Signal Processing Society Scholarship and the American Association of University Women (AAUW) International Fellowship.

\section*{Ethics Statement}

CASS has emerging applications in improving content accessibility and internationalization. Within a production context, CASS was developed to \textit{assist} human creatives with more tedious and repetitive aspects of cinematic audio production, in the hope that this would enable them to focus more of their time on the creative aspects of their work.

Divide and Remaster version 3 was developed using data derived exclusively from sources with explicit and permissive licenses, permitting commercial and derivative use. 
The dataset was developed in consultation with the lead authors of the original dataset and released under the same name with a major version change for continuity. 
The Bandit model employed is non-generative in nature; it can only filter audio content from its input and is incapable of outputting any audio content that was not already present within its input.


\bibliographystyle{IEEEtran}
\footnotesize
\bibliography{ctrl,references}

\end{document}